\documentclass[conference]{IEEEtran}
\IEEEoverridecommandlockouts
\usepackage{cite}
\usepackage{amsmath,amssymb,amsfonts}
\usepackage{algorithm}
\usepackage{algpseudocode}
\usepackage{graphicx}
\usepackage{textcomp}
\usepackage{xcolor}
\usepackage{glossaries}
\usepackage{multirow}
\usepackage{url}
\def\BibTeX{{\rm B\kern-.05em{\sc i\kern-.025em b}\kern-.08em
    T\kern-.1667em\lower.7ex\hbox{E}\kern-.125emX}}

\graphicspath{{figs/}} 

\usepackage{listings}

\lstdefinestyle{python}{
    language=Python,
    morekeywords={yield,List,Iterator}
}

\lstdefinestyle{c++}{
    language=c++,
    morekeywords={__shared__,__global__,if,for,int,double,__syncthreads()}
}

\usepackage{soul}

\newcommand{\indexadd}{\texttt{index\_add} }
\newcommand{\scatterreduce}{\texttt{scatter\_reduce} }
\newcommand{\atomicadd}{\texttt{atomicAdd} }
\newcommand{\threadfence}{\texttt{\_\_threadfence} }
\newcommand{\syncthreads}{\texttt{\_\_syncthreads} }

\begin{document}
\newacronym{HPC}{HPC}{high-performance computing}
\newacronym{DL}{DL}{deep learning}
\newacronym{AI}{AI}{artificial intelligence}
\newacronym{DOE}{DOE}{US Department of Energy}
\newacronym{DNN}{DNN}{deep neural network}
\newacronym{DNNs}{DNNs}{deep neural networks}
\newacronym{DFT}{DFT}{density functional theory}
\newacronym{FPNA}{FPNA}{floating-point non-associativity}
\newacronym{GPU}{GPU}{graphics processing unit}
\newacronym{FP}{FP}{floating-point numbers}
\newacronym{FP64}{FP64}{double precision floating point numbers}
\newacronym{FP32}{FP32}{single precision floating point numbers}
\newacronym{PDF}{PDF}{probability density function}
\newacronym{SPSA}{SPA}{simple-pass-with \texttt{atomicAdd}}
\newacronym{AO}{AO}{\texttt{atomicAdd}-only}
\newacronym{CU}{CU}{sum function from the CUB/HIPCUB library}
\newacronym{SPTR}{SPTR}{single-pass-with-tree-reduction}
\newacronym{FIFO}{FIFO}{first in first out queue}
\newacronym{TPRC}{TPRC}{two-passes-with-final-reduction-on-CPU}
\newacronym{SPRG}{SPRG}{single-pass-with-final-recursive-sum-on-GPU}
\newacronym{KL}{KL}{Kullback–Leibler divergence criterion}
\newacronym{GNN}{GNN}{graph neural network}
\title{Impacts of floating-point non-associativity on reproducibility for HPC and deep learning applications
\thanks{This manuscript has been authored by UT-Battelle, LLC under Contract No. DE-AC05-00OR22725 with the U.S. Department of Energy. The United States Government retains and the publisher, by accepting the article for publication, acknowledges that  the  United  States  Government  retains a non-exclusive, paid-up, irrevocable, world-wide license to publish or reproduce the published form of this manuscript, or allow others to do so, for United States Government purposes. The Department of Energy will provide public access to these results of federally sponsored research in accordance with the DOE Public Access Plan (http://energy.gov/ downloads/doe-public-access-plan).
\\
\\
$^\parallel$Equal contributions}
}
\author{
    \IEEEauthorblockN{
    Sanjif Shanmugavelu$^\parallel$\IEEEauthorrefmark{4}$^*$, 
    Mathieu Taillefumier$^\parallel$\IEEEauthorrefmark{2}$^*$, 
    Christopher Culver\IEEEauthorrefmark{4}, 
    Oscar Hernandez\IEEEauthorrefmark{3},
    Mark Coletti\IEEEauthorrefmark{3} and \\
    Ada Sedova$^\parallel$$^*$\IEEEauthorrefmark{3}}
    \IEEEauthorblockA{\IEEEauthorrefmark{4}Maxeler Technologies, a Groq Company. 3 Hammersmith Grove, W6 0ND, London, UK\\}
    \IEEEauthorblockA{\IEEEauthorrefmark{2}ETH Zurich/Swiss National Supercomputing Centre (CSCS). Via Trevano 131, 6900 Lugano, Switzerland}
    \IEEEauthorblockA{\IEEEauthorrefmark{3}Oak Ridge National Laboratory. Oak Ridge, TN, USA
    \\
    $^*$Corresponding emails:
    sshanmugavelu@groq.com
    tmathieu@ethz.ch
    sedovaaa@ornl.gov \\
    }
    }\vspace{-1em}


\maketitle

\begin{abstract}
Run to run variability in parallel programs caused by floating-point non-associativity has been known to significantly affect reproducibility in iterative algorithms, due to accumulating errors. Non-reproducibility can critically affect the efficiency and effectiveness of correctness testing for stochastic programs. Recently, the sensitivity of deep learning training and inference pipelines to floating-point non-associativity has been found to sometimes be extreme. It can prevent certification for commercial applications, accurate assessment of robustness and sensitivity, and bug detection. New approaches in scientific computing applications have coupled deep learning models with high-performance computing, leading to an aggravation of debugging and testing challenges. Here we perform an investigation of the statistical properties of floating-point non-associativity within modern parallel programming models, and analyze performance and productivity impacts of replacing atomic operations with deterministic alternatives on GPUs. We examine the recently-added deterministic options in PyTorch within the context of GPU deployment for deep learning, uncovering and quantifying the impacts of input parameters triggering run to run variability and reporting on the reliability and completeness of the documentation. Finally, we evaluate the strategy of exploiting automatic determinism that could be provided by deterministic hardware, using the Groq $\text{LPU}^{\text{TM}}$ accelerator for inference portions of the deep learning pipeline. We demonstrate the benefits that a hardware-based strategy can provide within reproducibility and correctness efforts.
\end{abstract}


\section{Introduction}
Run to run variability of a program, despite identical inputs and software stack, is usually assumed to be negligible, even in heterogeneous parallel programming \cite{lin2008principles}. However, variability caused by \gls{FPNA} coupled with asynchronous parallel operations such as reductions can be substantial \cite{Ahrens2020}, especially for massively parallel programs using iterative stochastic routines, such as those implementing optimization algorithms like conjugate gradient \cite{villa2009effects}. It can also mask errors within threshold-based correctness testing schemes \cite{thavappiragasam2022portability}, which are often used in scientific computing programs such as molecular simulation \cite{CP2K2020,salomon2013overview,legrand2020gpu}, making debugging difficult. \Gls{DNN} training also involves iterative, stochastic algorithms coupled with non-linear activation functions. This combination has been found to cause extreme sensitivity to bit-level numerical changes, such as those caused by \gls{FPNA} \cite{d9m,nagarajan2018impact,summers2021nondeterminism}. \gls{DNN} inference can also suffer from such sensitivity, due to the influence of non-linear activation functions, although the effects may be reduced due to the absence of the compounding errors caused by the iterative training schemes. Recently, a number of studies have shown that this run to run variability in the full \gls{DNN} training and inference pipeline can lead to unacceptably large differences in the predictions produced by a model. They have thwarted efforts to release reproducible commercial applications in safety-critical sectors such as medical diagnostics \cite{heumos2023mlf} and autonomous driving \cite{d9m}. In addition, recent reports have found that all of the major \gls{DL} software frameworks such as TensorFlow and PyTorch contain hundreds of bugs originating from all software levels, including environment misuse, incorrect assignment, and concurrency issues such as data races \cite{chen2023toward}; bugs in these frameworks can be disastrous \cite{jia2020empirical}, especially when they are silent \cite{tambon2024silent}, and high runtime variability can make it extremely difficult to detect bugs in these deep, multi-language, multi-level parallel stacks.

Within scientific \gls{HPC}, the incorporation of \gls{DL} into traditional approaches for simulation has become increasingly popular \cite{partee2022using,boyer2022scalable,wang2018deepmd}. Molecular simulation, for example, has used \gls{DNN} models for interatomic potentials, which promise quantum mechanical accuracy at the cost of simpler, empirical models; this translates to speedups of several orders of magnitude and the approach received the ACM Gordon Bell Award in 2020 \cite{jia2020pushing}. Our preliminary tests using identical inputs for training and inference pipelines for these types of models revealed a level of non-reproducibility for prediction of forces that would be unacceptable for traditional quantum mechanical \gls{HPC} programs \cite{APS-MM-24}. Besides non-deterministic kernels within the deep learning framework itself, any external kernels programmed by the scientific simulation developers which contain non-determinism, that feed into training pipelines \cite{jia2020pushing}, can also introduce large runtime non-reproducibility.

Here we present a systematic analysis of the effects of \gls{FPNA} within asynchronous parallel kernels on \gls{GPU} accelerators, together with several metrics that quantify this variability, within parallel programming schemes and in PyTorch functions, as well as within full end-to-end training and inference pipelines. We assess the impact of these effects on the ability to perform correctness testing and to produce reproducible scientific results. We also study solutions, using programming approaches, and via deterministic hardware for which we perform experiments with the Groq$\text{LPU}^{\text{TM}}$ deterministic inference chip. 

\section{Metrics for measuring the variability of non-deterministic functions}

We defined metrics $V$ for quantifying the run to run variability in output values for implementations of functions with scalar and multidimensional (array) inputs and outputs, following a similar approach used in error analysis \cite{Higham2002}. These are explained below. 
In all cases, $V = 0$ if two implementations are bitwise identical, nonzero otherwise, and increasing as variability increases. 

\subsubsection{Scalar-valued outputs}
We use $V_s(f) = 1 - \left|f_{\textsc{nd}} / f_{\textsc{d}}\right|$ to quantify the bitwise non-determinism between the outputs of two implementations of some function $f$,
where the subscripts $\textsc{nd}$ and $\textsc{d}$ label the non-deterministic and deterministic implementations respectively.

\subsubsection{Array outputs}
To quantify the bitwise variability of a non-deterministic implementation of a function producing a multidimensional array output, we define two different metrics. Let two implementations of a function $f$ be $f_1$ and $f_2$, which produce as outputs the arrays $\mathbf{A}$ and $\mathbf{B}$, respectively, each with dimensions $d_1, d_2, \ldots, d_k$ and $D$ total elements. The first metric is the elementwise relative mean absolute variation (ERMV),
\begin{equation}
V_{\text{ermv}}(f) = \frac{1}{D} \sum_{i_1=1}^{d_1} \cdots \sum_{i_k=1}^{d_k} \frac{|A_{i_1, \ldots, i_k} - B_{i_1, \ldots, i_k}|}{|A_{i_1, \ldots, i_k}|}.
\end{equation}
The second metric, the count (C) variability, measures how many elements in the multidimensional array are different, 
\begin{equation}
V_c(f) = \frac{1}{D} 
\sum_{i_1=1}^{d_1} \cdots \sum_{i_k=1}^{d_k} \mathbf{1}(A_{i_1, i_2, \ldots, i_k} \ne B_{i_1, i_2, \ldots, i_k})
\end{equation}
where $\mathbf{1}(\cdot)$ is the indicator function, which is 1 if the condition inside the parentheses is true, and 0 otherwise. 
Each of these metrics is zero if and only if the two multidimensional arrays $\mathbf{A}$ and $\mathbf{B}$ are bitwise identical. The count variability $V_c$ informs us of the percentage of varying array elements between the two output arrays, while the $V_{\text{ermv}}$ produces a global metric for the variability of the array outputs.

\section{Programming deterministic parallel sums}\label{sec:atomic_ops_for_sum}
In this section we first introduce \gls{FPNA} impacts for the parallel sum, then demonstrate several deterministic programming solutions, for GPUs using CUDA, and also using OpenMP for CPU or GPU. We then examine the performance impacts of these solutions, as well as programming productivity considerations. Parallel sums, as we will show in the subsequent sections, contribute some of the most substantial sources of variability to PyTorch functions on GPUs, and are thus responsible for concerning non-determinism in real-world \gls{DL} applications such as \gls{GNN}s, which make heavy use of them. A concerning connection to molecular physics and chemistry applications in scientific computing is the increased popularity of \gls{GNN}s for surrogate models in these applications \cite{batzner20223,kovacs2023evaluation}, especially considering the stringent precision requirements for these applications, as discussed below. 

Consider the sum $S_{\textsc{d}} = \sum_{i}^n x_i$ that we evaluate with a deterministic algorithm (the summation order is always the same). The simplest implementation adds the \gls{FP} $x_i$ in serial in the order they are stored. When parallelized with asynchronous operations and executed in an unspecified order, this is equivalent to applying a random permutation $P$ to the series $x_i$ before computing the serial sum, $S_{\textsc{nd}} = \sum_{i}^n x_{P(i)}$. 
\begin{table}[h]\vspace{-2mm}
\caption{Effects of permutations on sums of floating-point numbers}
\label{results:simple:python:code}
\centering
\begin{tabular}{|l|r|r|}
\hline
\bfseries size &  $\mathbf{S_{\textrm{nd}} - S_d}$ & $\mathbf{V_s}$\\
\hline\hline
100  &$-1.776356839400250 \text{e}^{-15}$ &$-2.220446049250313\text{e}^{-16}$\\
1000 &$1.776356839400250\text{e}^{-15}$ &$-6.661338147750939\text{e}^{-16}$\\
1000 &$5.329070518200751\text{e}^{-15}$ &$-1.776356839400250\text{e}^{-15}$\\
10000 &$-2.486899575160351\text{e}^{-14}$ &$1.110223024625157\text{e}^{-16}$\\
10000 &$3.197442310920451\text{e}^{-14}$ &$8.881784197001252\text{e}^{-16}$\\ 
100000 &$2.842170943040401\text{e}^{-13}$ &$1.110223024625157\text{e}^{-16}$\\
100000 &$5.684341886080801\text{e}^{-14}$ &$1.110223024625157\text{e}^{-16}$\\
1000000 &$4.263256414560601\text{e}^{-13}$ &$3.108624468950438\text{e}^{-15}$\\
1000000 &$1.705302565824240\text{e}^{-13}$ &$2.220446049250313\text{e}^{-15}$\\
\hline
\end{tabular}
\end{table}
To demonstrate the magnitude of the variability, we can generate lists $x_i$ of \gls{FP64} numbers of various lengths using Python, with, for example, $x_i\in \mathcal{N}(0, 1)$ the normal distribution of zero mean and $\sigma = 1$, and computed $S_\textsc{d}$ and $S_{\textsc{nd}}$ before and after applying a random permutation $P$ to $x_i$, repeating  ten times; Table~\ref{results:simple:python:code} shows the results. The variability can be larger than the tolerance of some of the correctness tests included in computational physics and chemistry programs; this same result obtains with a Boltzmann (exponential) distribution as well, which is the expected distribution for such calculations. For example, the quantum mechanics simulation program CP2K\cite{CP2K2020} uses a tolerance-based correctness testing scheme, and some of the tests have tolerances as low as $10^{-14}$ for values such as energy. This illustrates the problems that \gls{FPNA} can introduce in correctness testing schemes for computational tasks with stringent accuracy requirements. In addition, such programs often employ iterative solvers such as conjugate gradient, leading to accumulating \gls{FP} errors which can approach or exceed 20\% after six or seven iterations using double precision, as has been shown for \gls{HPC} systems such as massively multithreaded machines like the Cray XMT \cite{villa2009effects}.



\subsection{Examples of deterministic parallel sum implementations on GPUs with CUDA/HIP}
A difficulty for programming on GPUs is the absence of a global synchronization barrier at the kernel level.
To avoid races one can use (i) a single \texttt{atomicAdd} function, (ii) the implicit global synchronization happening when multiple kernels are added to the same stream 
(iii) one GPU thread to calculate the sum, (iv) the \threadfence instruction. 
\begin{lstlisting}[
   % float=h,
    belowskip=-1mm,
    language=c++,
    label=sum:single:thread,
    caption={Implementation of parallel sum using the \atomicadd \\instruction, and deterministic implementation using pairwise algorithm.}]

// implementation (AO) of the sum
__global__ void reduce_atomic_only(double *sdata, int size, double *res) {
    const int tid = threadIdx.x + blockIdx.x * blodkDim.x;
    if (tid >= size__) return;
    atomicAdd(res, sdata[tid]);
}

__device__ int retirementCount = 0;
// Implementation of the SPRG method 
__global__ void reduce_sprg(double *sdata, int size, double *res) {
    int tid = threadIdx.x + blockIdx.x * blodkDim.x;
    extern __shared__ double smem[];

    if (tid >= size) smem[threadIdx.x] = 0.0;
    else smem[threadIdx.x] = sdata[tid];
    __syncthreads();

    for (int offset = blockDim.x / 2; offset > 0; offset  /= 2) {
        if (threadIdx.x < offset) 
            smem[threadIdx.x] += smem[threadIdx.x + offset];
        __syncthreads();
    }

    if (threadIdx.x == 0) res[blockIdx.x] = smem[0];    
    __threadfences();

    bool __shared__ amLast = false;
    if (threadIdx.x == 0) {
        int prev = atomicInc(&retirementCount, gridDim.x);
        amLast = (prev == (gridDim.x - 1));
    }
    __syncthreads();

    if ((amLast) && (threadIdx.x == 0)) {
        for (int i = 1; i < gridDim.x; i++) 
            res[0] += res[i];
    }
}
\end{lstlisting}
Option (iii) is deterministic, but slow, as only one GPU thread computes the reduction. A code snippet can be found in the GitHub repository associated with this paper. Option (i) is only available for AMD GPUs (with HIP) in an unsafe compiler mode and is not considered for AMD GPUs here. This \gls{AO} method is
 shown in Listing~\ref{sum:single:thread} for the CUDA version. This approach is the easiest to program (only a few lines of code), but is actually sequential, and yet the instruction order is runtime dependent, making it non-deterministic. 
Option (ii) and (iv) use a pairwise algorithm \cite{Higham2002,Blanchard2020} coupled to data blocking for distributing the data over the different thread blocks and specifying the order for summation. 
Here, each thread of a given block adds 
 the elements $x_i$ 
 in pairs $t_{i} = x_{i} + x_{i + n/2}$, $i=1, \cdots,n/2$, done in parallel on GPU. This process is repeated  $log_2(n)$ times on $t_i$.
 We use the \syncthreads instruction after each step of the partial reduction for thread synchronization within the thread block. Shared memory is also used to improve data locality and performance. Partial results are then stored back in global memory.
We tested three different methods for accumulating these partial results. The \gls{SPSA}, uses an \atomicadd instruction to accumulate all partial sums instead of storing the results back in memory. 
This is again a simple solution from the programmatic perspective, but the implementation is not deterministic. The \gls{TPRC} method uses the property that two kernels launched on the same stream will execute sequentially following the submission order, introducing a synchronization barrier between them, or between a kernel and a data transfer between GPU and CPU. 
We choose the data transfer, and compute the final sum on CPU using the sequential recursive method. \gls{TPRC} is deterministic, but more sensitive to compiler optimizations because of vectorization.
The last approach 
uses an integer counter to keep track of the completed thread blocks. The \gls{CU} uses a similar technique. 
Each thread block increments this variable atomically before exiting. The last block incrementing the counter is responsible for the remaining reduction, and can be performed in two ways. The \gls{SPTR} method uses the same block reduction algorithm than for the first stage while the \gls{SPRG} variant (see listing~\ref{sum:single:thread}) 
uses the recursive sum. 
\gls{SPTR} and \gls{SPRG} are both deterministic by construction. \gls{SPRG} and \gls{SPTR} are only possible if we use the \threadfence instruction to avoid data race conditions. This instruction ensures that all writes issued by the calling thread are finished before the calling thread runs the next instruction, and notifies all the other running threads that the memory operation is over. The instruction is not a global synchronization barrier. 
\begin{table}[h]
    \centering
    \caption{Different implementations of the parallel sum in CUDA. Non-deterministic implementations shown in red.}
    \label{tab:sum:implementations}
    \begin{tabular}{|l|l|l|l|}
    \hline
    Method &deterministic &\# of kernels &synchronization methods\\
    \hline\hline
    \textcolor{olive}{\gls{CU}} & Yes & - & \threadfence\\
    \textcolor{olive}{\gls{SPTR}} & Yes &1 & \threadfence\\
    \textcolor{olive}{\gls{SPRG}} &Yes &1 &\threadfence\\
    \textcolor{olive}{\gls{TPRC}} &Yes &2 &stream synchronization \\
    \textcolor{red}{\gls{SPSA}} &No &1 &\atomicadd\\
    \textcolor{red}{\gls{AO}} &No &1 &\atomicadd\\
    \hline
    \end{tabular}
    \vspace{-2mm}
\end{table}
A summary of the main properties of each implementation is provided in Table~\ref{tab:sum:implementations} while codes snippets and full reference codes are provided in the GitHub repository~\cite{correctness-github}.

\subsection{Other approaches: Parallel sums with OpenMP}
OpenMP, a directive-based API for shared memory and accelerator parallel programming, supports data reduction to perform parallel calculations that are portable across architectures. The OpenMP programming model uses reduction scoping clauses to specify the regions of code where the reduction takes place. These regions can include parallel fork-join, tasks, sections, SIMD, and scope constructs~\cite{openmp52}, which can be executed on a host or a device such as a GPU. The OpenMP specification does not specify the location and ordering in which the values are combined, thus, 
bitwise determinism is not guaranteed. 
As a result, implementing a deterministic parallel reduction in OpenMP will require using constructs that can enforce ordering when reducing private and shared variables. An ordered construct can define a structured block within a loop, SIMD, or loop SIMD region. This construct enforces sequential execution for a specified region of code according to the loop iterations, while enabling parallel execution for code outside the region. When the thread handling the first iteration of the loop reaches the ordered construct, any other thread that encounters the ordered construct in its iteration will wait at the start of the ordered region until all ordered regions from previous iterations have been executed. To order the entire loop, an ordered clause can be used.

\begin{lstlisting}[language=c, caption=An ordered parallel reduction implementation OpenMP, label=lst:ordered_reduction]
#pragma omp parallel for ordered &
        reduction(+:sum)
    for (i = 0; i < size; i++) {
       #pragma omp ordered
        {
            sum += array[i];
        }
    }
\end{lstlisting}

\begin{lstlisting}[language=c, caption=An ordered parallel reduction implementation OpenMP using \\
target construct, label=lst:target_ordered_reduction]
#pragma omp target parallel for simd ordered &
        reduction(+:sum)
    for (i = 0; i < size; i++) {
            sum += array[i];
        }
\end{lstlisting}

Listing~\ref{lst:ordered_reduction} shows the use of a block-ordered directive on a reduction. A similar effect can be achieved while using an ordered clause.  Listing~\ref{lst:target_ordered_reduction} shows how to use an ordered clause inside a target region that executes on a device or GPU.
When comparing the results of listing~\ref{lst:ordered_reduction} with a reduction without the ordered directive, using gcc 12.2.0 on CPU, we get the results shown in Table~\ref{tab:reductions}; the ordered version produces deterministic results.
\begin{table}[htbp]
\centering
\caption{Results of the normal and ordered reductions using OpenMP on CPU}
\label{tab:reductions}
\begin{tabular}{|l|c|c|}
\hline
\textbf{Trial} & \textbf{Normal Reduction} & \textbf{Ordered Reduction} \\ \hline \hline
1 & 2.3542548638889723e-07 & 2.3542548638889725e-07 \\ 
2 & 2.3542548638889731e-07 & 2.3542548638889725e-07 \\ 
3 & 2.3542548638889725e-07 & 2.3542548638889725e-07 \\ 
4 & 2.3542548638889723e-07 & 2.3542548638889725e-07 \\ 
5 & 2.3542548638889717e-07 & 2.3542548638889725e-07 \\ 
6 & 2.3542548638889725e-07 & 2.3542548638889725e-07 \\ 
7 & 2.3542548638889728e-07 & 2.3542548638889725e-07 \\ 
8 & 2.3542548638889728e-07 & 2.3542548638889725e-07 \\ 
9 & 2.3542548638889731e-07 & 2.3542548638889725e-07 \\ 
10 & 2.3542548638889728e-07 & 2.3542548638889725e-07 \\ \hline
\end{tabular}
\vspace{-1em}
\end{table}
We achieve this determinism by taking advantage of OpenMP static scheduling and the ordered clause and directive. However, the ordered directive is not available for the team distribute directive. This limitation is important for reductions that execute on a device, as the order of iterations between teams is not possible; a user-defined or lower-level reduction implementation as previously described will be needed.

\subsection{Statistical properties of the variability of non-deterministic parallel sums using CUDA or HIP on different GPUs}

\begin{figure}[htbp]
\centering
\includegraphics[width=0.38\textwidth]{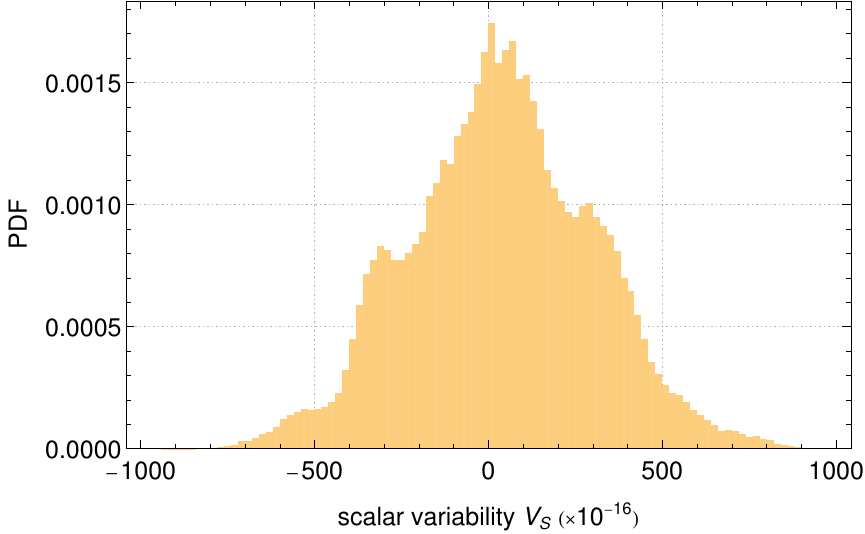}
\caption{\gls{PDF} of the scalar variability $V_s$ for 1 M FP64 numbers sampled from the uniform distribution $U(0, 10)$ when \gls{AO} is used for the nondeterministic implementation, on V100.}
\label{pdf:full:atomic:add}
\label{distribution:atomicadd:only}
\vspace{-1em}
\end{figure}

It is often assumed that the variability due to \gls{FPNA} can be described as Gaussian noise, but we could not find evidence supporting this in the literature. We therefore performed numerical experiments to estimate the \gls{PDF} of $V_s(S)$. 
We generated a set of 100 arrays of one million \gls{FP64} elements each taken from the uniform distribution $U(0, 10)$, applied \gls{SPSA} 10000 times to each array, and evaluated the variability $V_s$, using \gls{SPTR} for the deterministic summation, returning one million values for $V_s$. $S_\textrm{d}$ thus refers here to \gls{SPTR} while $S_{\textrm{nd}}$ refers to \gls{SPSA}. We then repeated the experiment, replacing the uniform distribution with a normal distribution of zero mean and standard deviation 1. 
\begin{figure}[htbp]
\centering
\includegraphics[width=0.45\textwidth]{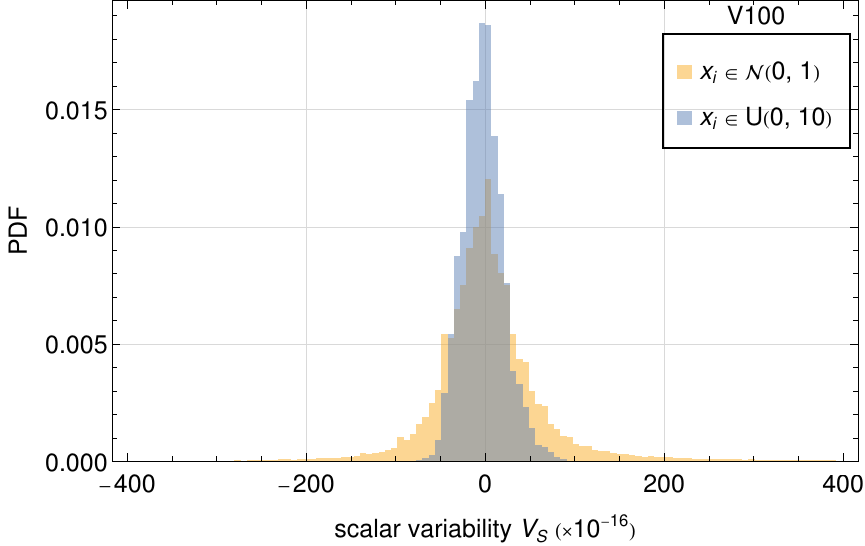}
\caption{Probability density of the variability $V_s$ for sums of 1 M \gls{FP64} numbers sampled from normal and uniform distributions on the V100 GPU. 
Kernel parameters are $N_t=64$ and $N_b =7813$ for both \gls{SPTR} and \gls{SPSA}.} 
\label{fig:distribution}
\vspace{-2mm}
\end{figure}
The resulting \gls{PDF}s are shown in Fig.\ref{fig:distribution} for both distributions for the V100; for the Mi250X and GH200 GPUs the results can be found in the GitHub repository~\cite{correctness-github}. The 
means and standard deviations of $V_s$ are different between the GPU types, while the shapes are similar. Using \gls{KL} analysis, we find that all \gls{PDF}s for \gls{SPSA} do converge towards a normal distribution whose mean and standard deviation depend on $x_i$ and the GPU family. However, when we repeated the experiments replacing \gls{SPSA} with \gls{AO} on the NVIDIA GPUs, using a sample size of 500000 sums (results shown in Fig.~\ref{pdf:full:atomic:add} for the V100), the distribution is found not to be normal, showing that the assumption of Gaussian noise is invalid in general. The reasons for this distribution are unclear, as the NVIDIA runtime scheduler details are proprietary. 

We also calculated the dependence $\text{Max} \left|V_s\right|$ as a function of the array size $n$ for the same sequences used for Fig.~\ref{fig:distribution}. The data can be fit with a power law of the form $\beta n^\alpha$. $\text{Max} \left|V_s\right|$ is proportional to $\sqrt{n}$ when $x_i\in U(0, 10)$. The exponent is larger for $x_i\in\mathcal{N}(0,  1)$, showing that the range of the numbers also plays a role. 



\subsection{Performance comparisons}

 To illustrate the performance impact of these different strategies, we measured the time needed to compute 100 sums of 4194304 \gls{FP64} elements taken from the uniform distribution, using \gls{SPTR}, \gls{TPRC}, \gls{CU}, \gls{AO} and \gls{SPSA}. The only nondeterministic algorithms in this tests are \gls{AO} and \gls{SPSA}. We compute the negative quantity $P_s = 100 (1 - t_i)/\text{min}(t_i)$ where $\text{min}(t_i)$ is the minimum of all timings $t_i$, to compare the performance of all implementations to the fastest one. The other parameters controlling the algorithm are the thread block size and the number of thread blocks. 
 Details on the choice of these settings are provided in the associated repository~\cite{correctness-github}.

The main results are summarized on Tab.~\ref{table:timings}. We find that \gls{AO} is 2 orders of magnitude slower than the fastest implementations. The fastest sum implementation depends on the GPU version. The non-deterministic  \gls{SPSA} seems to be favored on NVIDIA GPU, followed by \gls{SPTR} and \gls{CU}. The difference in timings between \gls{SPSA} and \gls{SPTR} is less than 0.2\% for the V100 GPU but can reach up to 7.8\% on the GH200. \gls{CU} 
suffers a 4.5\% penalty on GH200. Most implementations on V100 are within 0.5\% to 1\% of each other. 
The performance penalty on GH200 is more spread than on V100.
\gls{TPRC} is the fastest implementation on the Mi250X GPU followed by \gls{CU}. 
\begin{table*}
\centering
\caption{Timing and performance penalty of parallel sum implementations on different GPUs for 100 sums of 4194304 \gls{FP64} numbers and varying kernel parameters. Timings averaged over 10 consecutive runs; non-deterministic algorithms indicated in red. Standard deviations in parentheses.}
\label{table:timings}
\begin{tabular}{|c|c|c|c|c|}
\hline
GPU &implementation &($N_t\times N_b$) &time for 100 sums (in ms) &$P_s$ (in \%) \\
\hline\hline
&\textcolor{red}{\gls{SPSA}} & $(512  \times 128)$  &$6.456(0.008)$   &$0$\\  
&\textcolor{olive}{\gls{SPTR}}  & $(512  \times 128)$  &$6.469(0.011)$   &$-0.198538$\\  
V100 &\textcolor{olive}{\gls{TPRC}}  & $(512  \times 128)$  &$6.491(0.006)$   &$-0.528162$\\
&\textcolor{olive}{\gls{CU}} & (unknown) &$6.87697$ &$-6.51331$ \\
&\textcolor{red}{\gls{AO}} & (fixed parameters) &$872.004(0.022)$ &$-28781.3$\\
\hline
\hline
&\textcolor{red}{\gls{SPSA}} &$(512  \times 512)$  &$3.019(0.022)$   &$0$\\
&\textcolor{olive}{\gls{CU}}  &(unknown)      &$3.1553$    &$-4.50533$\\
&\textcolor{olive}{\gls{TPRC}}  &$(512 \times 512)$ &$3.226(0.030)$ &$-6.87139$\\
GH200 &\textcolor{olive}{\gls{SPTR}} &$(512 \times 512)$ &$3.254(0.014)$ &$-7.79916$ \\
&\textcolor{red}{\gls{AO}} &(fixed parameters) &$738.687(0.037)$ &$-24365.7$\\
\hline
\hline
&\textcolor{olive}{\gls{TPRC}}  &$(512  \times 256)$  &$6.275(0.012)$   &$0$\\  
&\textcolor{olive}{\gls{CU}}  &(unknown)    &$6.378$   &$-1.63577$\\  
Mi250X &\textcolor{red}{\gls{SPSA}} &$(512\times 256)$ &$6.394(0.032)$ &$-1.90101$\\
&\textcolor{olive}{\gls{SPTR}} &$(256\times 512)$ &$6.552(0.023)$ &$-4.4171$\\
\hline
\end{tabular}
\vspace{-1em}
\end{table*}
Our results show that the performance of a deterministic implementation can be faster or 
 only slightly slower than its non-deterministic counterpart, depending on the GPU type. While timings can also change depending on the GPU load,
these results suggest that there is no reason to calculate a parallel sum using nondeterministic \atomicadd operations, as the performance benefit is marginal at best. 


\section{Non-determinism in PyTorch Functions}\vspace{-1mm}

As discussed in the introduction, \gls{FPNA} can lead to large variations in identical training and inference pipelines for deep learning, due to compounding effects within training, along with the impact of nonlinear activation functions. Analogously to the above section, we explore \gls{FPNA}-induced variability for the PyTorch operations that are documented to have non-deterministic behavior~\cite{PyTorch-Non-Determinism}. 
PyTorch makes it easy for end users to construct neural networks out of high-level modules and deploy software on devices such as GPUs without the need for direct GPU programming. GPU kernels are provided by vendor libraries such as NVIDIA's cuDNN and AMD's MIOpen, and sometimes by PyTorch developers. As demonstrated in the previous section, any kernels PyTorch uses that rely on atomic operations will not be deterministic, leading to output variability. Furthermore, to be hardware agnostic and computationally efficient there are cases where strategies have been developed to determine the optimal computational kernel at runtime, also causing non-determinism.  

Apart from the choice of kernel at runtime and atomic operations, there are several other ways in which PyTorch may induce variability. These include having an unset random number generator seed, having unset CUDA environment variables, uninitialized GPU memory and communication between devices. To focus exclusively on variability emergent from the first two sources, we a single GPU and remove those other sources of variability.   
We control whether or not non-deterministic kernels are available through the environment variable that PyTorch provides~\cite{PyTorch-Non-Determinism}. We note that this documentation may not necessarily be completely accurate across all systems and software versions, as we received a runtime error when trying to obtain a deterministic result for \texttt{scatter\_reduce}, suggesting the potential difficulties high level users may experience when trying to control determinism within this deep software stack.

We performed experiments in two ways, depending on whether or not a deterministic kernel exists. If there is a deterministic option, then the output multidimensional array $\mathbf{A}$ is fixed by that implementation.  We compare it with the tensors $B_i$ with $i$ labelling $N$ runs of the non-deterministic implementation. If there is no non-deterministic kernel, we choose the first non-deterministic invocation as a reference, $\mathbf{A}=\mathbf{B}_0$.  
For each experiment we also present results for measurements of kernel runtime for non-deterministic implementations on the GPU, and deterministic implementations on the GPU and LPU architectures, to expose the performance costs of using deterministic operations. Runtime measurements are only for the execution time of the kernel, excluding data transfers to/from the device. For the GPU we make many measurements to report the average and standard deviation of the execution time. On the LPU architecture the runtime for the PyTorch function is reported as a fixed number since the cycle-by-cycle execution is determined ahead of time~\cite{groq-tsp}.

We performed a sweep over operation hyperparameters using 10000 runs on an H100 and observed only a handful of functions specified in the PyTorch documentation\cite{PyTorch-Non-Determinism} to exhibit non-determinism. We list these operations in Table \ref{tab:Operations} alongside minimum and maximum $V_{\text{ermv}}$. We refer the reader to our repository~\cite{correctness-github} for in-depth details on the full hyperparameter sweep, where, for example, we consider kernel size, stride and padding when testing convolution operations. 

Our test results indicated that the main factor in ability to observe output variability in these kernels is the size of the input tensor, which gives more chances to observe the switching of order at runtime. For functions which perform reductions on an input tensor, we found that the ratio of the output dimension to input dimension is another key factor. We explore this further, restricting our analysis to \scatterreduce and \texttt{index\_add}, which have such reductions.\vspace{-1mm}

\begin{table}[h]
    \centering
    \caption{Max and min variability for non-deterministic pytorch operations over all hyperparameters tested.}
    \label{tab:Operations}
    \begin{tabular}{|l|l|l|}
        \hline
        \textbf{Operation} & \textbf{min$(V_{\text{ermv}}) / 10^{-7}$} & \textbf{max$(V_{\text{ermv}}) / 10^{-6}$} \\
        \hline
        \hline
        ConvTranspose1d & 1.52 & 2.36\\ 
        \hline
        ConvTranspose2d & 1.29 & 4.52\\ 
        \hline
        ConvTranspose3d & 0 & 1.34\\
        \hline
        cumsum & 0 & 0.52 \\
        \hline
        index\_add & 0 & 5.03 \\
        \hline
        index\_copy & 0.36 & 2.08 \\
        \hline
        index\_put & 0 & 1.68 \\
        \hline
        scatter & 0 & 3.82 \\
        \hline
        scatter\_reduce & 0 & 3.35 \\
        \hline
    \end{tabular}
    \vspace{-1em}
\end{table}

\subsection{Case studies of PyTorch kernels}\vspace{-1mm}

The \scatterreduce function updates an output array $Y$ by applying a reduction operation on values from the input array $X$, according to indices specified in an index array $I$, for example, when reducing a one dimensional array with summation reduction, $Y[i] =  \sum_{j} \left( \left\{ X[j] \mid j = I[i] \right\} \right)$. This can be generalized to arrays with arbitrary dimensions and different reduction functions. The \indexadd operation updates the output array $Y$ by adding values from an input array $X$ according to the indices specified in an index array $I$.  As an example, for two dimensional input and output arrays with a summation reduction over the first dimension, the elements of $Y$ are
 $ Y[i, j] = Y[i, j] + X[k, j],  \text{ for} \ i = I[k], \forall k $. This can similarly be generalized to arbitrary dimensions and different reductions other than addition. Both of these operations reduce the size of an input tensor, so we define the reduction ratio $R$ as the ratio of the output dimension size to source dimension size. When $R=1$, the arrays have the same size, and smaller values of $R$ correspond to larger reductions of the source array down to at most a singular value, along the axis of reduction. For the index tensors of both operations, we use random integers drawn from a uniform distribution to choose arbitrary values from the source tensor. The motivation for this is to mimic an arbitrary graph structure, where the reduction is happening over nodes that share an edge, although in this case we do not \emph{a priori} expect any specific structure or hierarchy to be emergent.

\begin{figure}[htbp]
\includegraphics[width=0.48\textwidth]{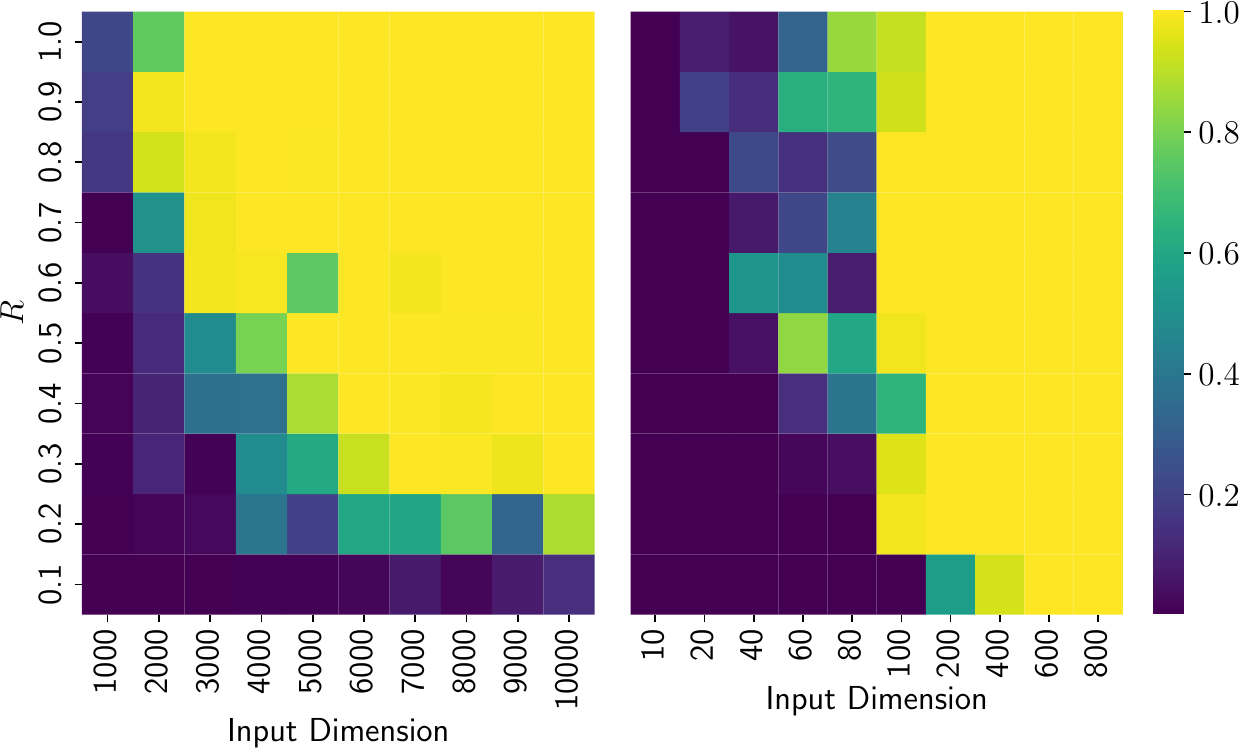}
\caption{Heatmaps of count variability ($V_c$) per run/iteration for 1,000 runs of the non-deterministic implementation of \scatterreduce~(left) and \indexadd~(right) for different reduction ratios and input dimensions. Note the input dimension for \indexadd is two dimensional square arrays, while the input dimension for \scatterreduce is one dimensional.}
\label{fig:pytorch_heatmaps}
\vspace{-1mm}
\end{figure}

Fig.~\ref{fig:pytorch_heatmaps} shows heat maps of $V_c$ as a function of $R$ and input dimension for the \scatterreduce and \indexadd operations with summation reductions. There is a clear trend of increased variability as a function of each of these parameters. In the case of larger input dimension, there are more opportunities for runtime non-determinism to occur. For small reduction ratios there is less variation. We suspect this is due to the fact that our index tensor is random, meaning that elements being reduced will not be located locally in memory. For both operations we observe $V_c$ near one, meaning that most runs have a unique output. This is the worst case for reproducibility and error debugging.
\begin{figure}[htbp]
\includegraphics[width=0.48\textwidth]{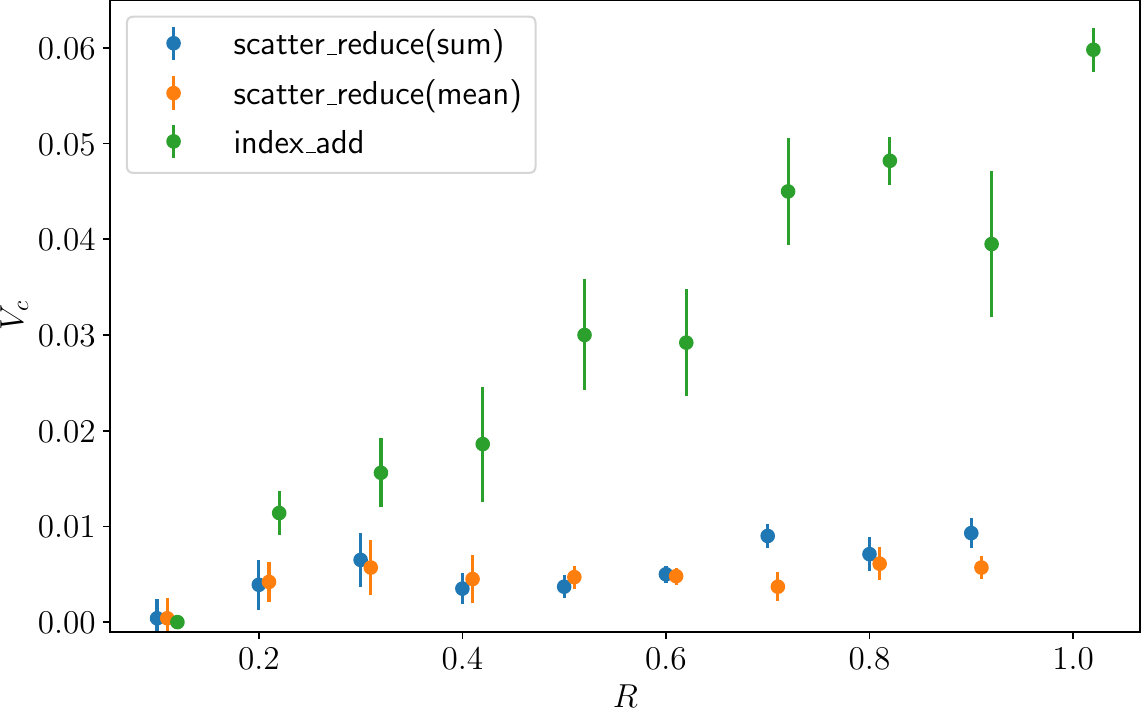}
\caption{Plot of the count variability for different reduction ratios for the scatter reduce and index add pytorch kernels.  For the scatter reduce kernel we use an array of 2,000 elements, while for the index add we use an array of 100 elements.}
\label{fig:variability_count}
\vspace{-2mm}
\end{figure}

\begin{figure}[htbp]
\includegraphics[width=0.48\textwidth]{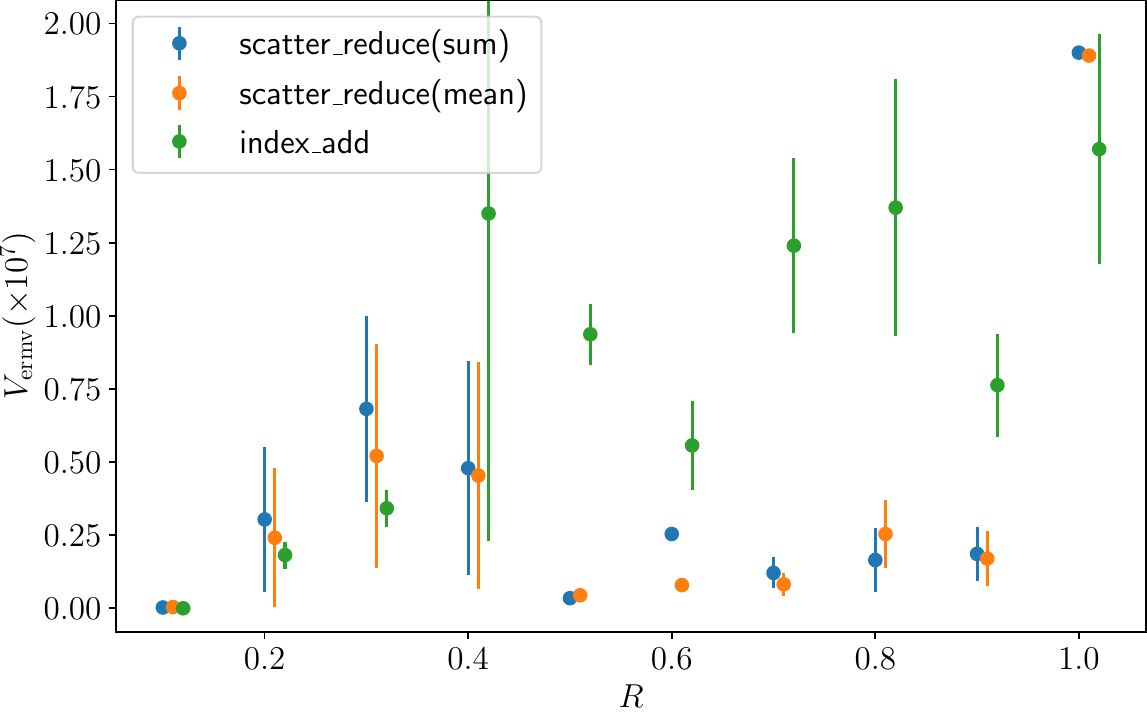}
\caption{Plot of the tensor variability for different reduction ratios for the scatter reduce and index add PyTorch kernels.  For the scatter reduce kernel we use an array of 2,000 elements, while for the index add we use an array of 100 elements.}
\label{fig:variability}
\vspace{-2mm}
\end{figure}

Figures~\ref{fig:variability_count} and~\ref{fig:variability} show $V_{\text{ermv}}$ and $V_c$ as a function of reduction ratio. For the \scatterreduce operation we use a one dimensional array with 2,000 elements, and for \indexadd we use a two dimensional square array with 100 $\times$ 100 elements.  These array sizes are selected to be in an interesting regime of reduction ratio behavior as observed from the heat-maps. We notice a fairly constant $V_c$ between  0.005 and 0.01 for \scatterreduce with both the \texttt{sum} and \texttt{mean} reductions. The $V_c$ for this operation at a reduction ratio of 1.0 has a value around 0.10 (not plotted) which is a significant jump. We notice similar behavior for $V_{\text{ermv}}$ for this operation. Note for the case where the output array has only one element, we recover the previous problem statement of computing the sum of an array, covered in Section \ref{sec:atomic_ops_for_sum}. For \texttt{index\_add}, $V_c$ is increasing almost linearly with reduction ratio. The errors are inconsistently sized across reduction ratios, indicating different behavior at each reduction ratio. We also notice an approximately linear trend between $V_{\text{ermv}}$ and reduction ratio, with a particularly large error bar at $R=0.4$. The lack of trend for the errors on the variability requires further analysis.
\vspace{-2mm}
\begin{table}[h!]
\centering
\caption{Average kernel runtime for \scatterreduce and \indexadd kernels on the H100 and Groq using deterministic~(D) and non-deterministic~(ND) implementations. Standard deviations in parentheses.}
\label{table:latency_operations}
\begin{tabular}{|c|c|c|c|}
\hline
Operation & Implementation & H100~$\left(\mu s\right)$ & Groq~$\left(\mu s\right)$ \\ \hline\hline
\scatterreduce & D & N/A & $10.5$ \\
\texttt{(sum)}               & ND & 30.2(1.4) & N/A \\ \hline 
\scatterreduce & D & N/A & 28.9 \\ 
 \texttt{(mean)}             & ND & 74.9(1.4) & N/A \\ \hline
\indexadd                    & D & 161(4) & 12.0 \\ 
                             & ND & 12.8(26) & N/A \\ \hline
%
%
\end{tabular}
\end{table}

We now investigate performance effects of deterministic and non-deterministic operations, performed on the H100 and LPU architectures. Up until now, the LPU architecture has been left out of the discussion of variability because its hardware operates deterministically~\cite{groq-tsp}. In Table~\ref{table:latency_operations} we report the average kernel runtime for the \scatterreduce with input dimension 1,000 and reduction ratio $0.5$ and \indexadd with input dimension 1,000 $\times$ 1,000 and reduction ratio $0.5$ on both the H100 and LPU architectures. For the latter we found the non-deterministic implementation on the H100 to be faster. This trend is not observed across all the operations in the documented list \cite{PyTorch-Non-Determinism}, however. For \scatterreduce and \indexadd the LPU architecture, which is by default deterministic, is faster than all GPU implementations.  
Runtimes for the complete set of functions in \cite{PyTorch-Non-Determinism} are given in the supplemental repository~\cite{correctness-github}. \vspace{-1mm}

\section{Effect of non-determinism on full deep learning workflows}

Non-determinism in PyTorch kernels can adversely affect  upstream tasks in model training and inference as found in real-world applications.  
To highlight this, we focus on training and inference for a Graph Sample and Aggregate~(GraphSAGE) neural network.\vspace{-1mm}

\subsection{GraphSAGE Convolution Network}

Graph Neural Networks (GNNs) are extensively used for analyzing graph-structured data, such as social networks and molecular structures. GNNs operate on the principle of message passing, where each node in the graph aggregates information from its neighbors to update its own representation.  The aggregation of information in software is often implemented via scatter and gather operations, which we have shown above have severe runtime variability. A layer in a GNN is defined by 
$h_v^{(k)} = \text{U}^{(k)}\left(h_v^{(k-1)}, \text{ A}^{(k)}\left(\{h_u^{(k-1)} \mid u \in \mathcal{N}(v)\}\right) \right) $
%
where \( h_v^{(k)} \) is the representation of node \( v \) at layer \( k \), \( \mathcal{N}(v) \) denotes the set of neighbors of node \( v \), and 
U and A are functions defining the update and aggregation steps, respectively. GraphSAGE is a popular graph convolution that uses
functions such as the sum and mean for the aggregation. Implementations often use the \scatterreduce and \indexadd operations discussed above, for example in the PyTorch Geometric~\cite{PyTorch-Geometric} library.  
\subsection{Results}

We trained a GNN with two SAGEConv layers on the Cora dataset. This dataset is widely used for GNN research and benchmarks; it consists of 2,708 scientific publications classified into one of seven classes. The graph structure is created with 5,429 links representing citations between these publications. Each publication is described by a 1,433-dimensional feature vector. 
Training of the SAGEConv model is performed on this data set with a 10 epoch training run. The only source of non-determinism in our implementation of this \gls{DNN} is the \indexadd operation.
We studied the variability of model weights over 10 epochs and found that mean $V_{\text{ermv}}$ increased from  from epoch 1 to 10, while the standard deviations also increased: 1.414(0.05) to 1.418(0.23).
This may indicate that, as expected, non-determinism results in a compounding increase in the variability with more kernel calls. At the end of the training loop, all 1,000 models had a unique set of model weights: $V_c~1$. Despite this  variability all models converge to similar loss values. Unlike stochastic training using random number generators for initialization, the resulting 1,000 models are completely non-reproducible, even for a single user on a single machine.\vspace{-1mm}

\begin{table}[h!]
\centering
\caption{$V_{\text{ermv}}$ and $V_c$ for different training and inference deterministic~(D) and non-deterministic~(ND) combinations on GPUs. Standard deviations in parentheses.}
\label{table:training_h100}
\begin{tabular}{|c|c|c|c|}
\hline
Training & Inference & $V_{\text{ermv}}/ 10^{-6}$ & $V_c$ \\ \hline\hline
D & D & 0(0) & 0(0) \\ \hline
              & ND & $2.63(0.12)$ & $0.12(0.2)$ \\ \hline 
ND & D & $4.27(0.21)$ & $0.16(0.2)$ \\ \hline 
                & ND & $5.08(0.23)$ & $0.21(0.4)$ \\ \hline
\end{tabular}
\end{table}



We also measured the different combinations of training and inference stages cumulatively, by creating 1,000 models under four conditions. These are deterministic training and deterministic inference, deterministic training and non-deterministic inference, non-deterministic training and deterministic inference, and lastly non-deterministic training and non-deterministic inference.  The $V_{\text{ermv}}$ and $V_c$ for each of these experiments is presented in Table~\ref{table:training_h100}. As expected, the non-deterministic training with non-deterministic inference has the most severe variations, and while training seems to incur more variability, inference variability contributes a non-negligible amount. The runtime for the GraphSAGE model training for the full 10 epochs is $0.48(1)$ seconds with a deterministic operation and $0.18(1)$ seconds with a non-deterministic one. The inference runtimes for a single input are given in Table~\ref{table:graphsage-inference}; the deterministic implementation is slower than the non-deterministic one on GPU. Inference on the LPU accelerator is 30 times faster than the fastest PyTorch implementation, consistent with results presented in \cite{ISC2023-GNN}. \vspace{-2.5mm}

\begin{table}[h]
    \centering
    \caption{H100 and Groq kernel runtime for deterministic and non deterministic inference of the GraphSAGE model. }
    \begin{tabular}{|c|c|c|}
        \hline
        Inference & H100~(ms) & Groq~(ms) \\ \hline\hline
        Deterministic & $3.92(0.2)$ & $0.066$ \\ \hline
        Non Deterministic & $2.17(0.1)$ & N/A \\ \hline
    \end{tabular}
    \label{table:graphsage-inference}
    \vspace{-2mm}
\end{table}\vspace{-2mm}

\section{Conclusions}\vspace{-1mm}
The analyses presented here highlight variability from non-deterministic kernels for basic parallel reductions, PyTorch functions, and resulting adverse effects on a full \gls{DL} training and inference pipeline. The variability from non-deterministic reductions can approach the tolerance thresholds used in high-accuracy molecular simulation correctness tests, and for \gls{DL}, will produce a unique and non-reproducible set of model weights with each run using identical inputs. These effects were pronounced for \scatterreduce and \indexadd operations, which are used extensively in GNNs. GNNs have become a key tool for \gls{DL} models used in molecular simulation. In terms of productivity and performance, using deterministic programming patterns may require more effort or experience, especially to maintain performance, and may not be attainable for all algorithms. Within large \gls{DL} framework stacks, providing deterministic solutions is left to the developers. We found that documentation on which functions in PyTorch are deterministic may be erroneous or incomplete, highlighting the large burden placed on end users in development of reproducible \gls{DL} workflows. The benefits of deterministic chip designs, therefore, may extend beyond performance improvements, to reproducibility facilitation and support, increasing productivity and correctness. The non-determinism studied here results from asynchronous atomic operations on GPUs, but in HPC and distributed settings there will also be inter-chip and inter-node communication, such as with MPI, leading to more runtime variation. On the LPU architecture, inter-chip communication can be software scheduled, removing such communication variations~\cite{groq-communication}. Future work should explore effects such as this in addition to the ones presented here. In addition, to mitigate the non-reproducibility produced by the training portions of a \gls{DL} workflow, deterministic hardware for training could also be proposed; emerging \gls{DL} chip designs may support such  solution and can be evaluated. \vspace{-2mm}
\section{Acknowledgements}\vspace{-1.5mm}
This work was supported in part by the ORNL AI LDRD Initiative and in part by Swiss Platform For Advanced Scientific Computing (PASC), and used resources of the OLCF, a DOE Office of Science User Facility [DE-AC05-00OR22725], and Swiss National Supercomputing Centre (CSCS).


\bibliographystyle{IEEEtran}
\bibliography{references,coletti,hardware,software}

\section*{Artifact Description}

All source code and inputs for all tests and programs reported in this paper can be found in our devoted GitHub repository at
\url{https://github.com/minnervva/fpna-robustness}. The main directories \texttt{codes/test\_reduce} and \texttt{codes/test-suite} contain the codes used in Section III and IV of the main text. Repository organization, installation, usage, and contributing instructions are documented in the README.

\subsection{Experimental Methods}
In the following we describe the hardware and systems software used to perform our numerical analyses and the mathematical definitions developed to quantify output variability caused by non-determinism due to \gls{FPNA} in parallel kernels, studied as isolated test cases and within the PyTorch deep learning framework high-level operations.

Tests on GH200 are run on the Alps supercomputing system located at CSCS running SLE 15 (enterprise). Each node has 4 GH200 with 72 ARM cores, 128 GB LPDDR5, and one H100 GPU with 96 GB HBM3 memory each. 

Tests for the V100 are run on the Summit supercomputer at the Oak Ridge Leadership Computing Facility (OLCF), running Redhat OS 8. Summit is an IBM system; each IBM Power System AC922 node has two Power9 CPUs with 512 GB of memory and 6 V100 NVIDIA GPU with 16GB of HBM2 memory. 

Tests on Mi250X AMD GPU are obtained on the Frontier supercomputer at OLCF, running SLE 15 (enterprise). Frontier is an HPE Cray EX supercomputer; each Frontier compute node has a 64-core AMD ``Optimized 3rd Gen EPYC” CPU with 512 GB of DDR4 memory and 4 AMD MI250X GPUs, each with 2 Graphics Compute Dies (GCDs) for a total of 8 GCDs per node. 

Tests on the NVIDIA H100 are performed on a Groq host node3 in Groq ZankerLab running Ubuntu LTS 22.04.06. This machine has 2 H100's with 40GB HBM3 memory each and has an AMD EPYC 7302 16-Core Processor host CPU with 2 sockets, 16 cores per socket and 2 threads per core. Tests on the LPU accelerator are run on the $\text{GroqNode}^{\text{TM}}$ server r01-gn-01 in Groq ZankerLab with $\text{GroqWare}^{\text{TM}}$ SDK 0.11 and Ubuntu LTS 22.04. A GroqNode has 8 Groq LPUs connected with a fully connected 88 RealScale™ chip-to-chip connectors and 2 AMD EPYC 7313 processors (3GHz, 16C/32T, 155W TDP each).
\end{document}